\documentclass[final,3p]{elsarticle}
\usepackage{amsmath,amssymb,graphicx}
\usepackage{amsfonts,amsthm}

\usepackage{amsmath,amssymb,graphicx,amsthm}
\usepackage{color}
\usepackage{soul}

\begin{document}

\begin{frontmatter}

\title{Grasping Complexity}
\author{A.~N.~Gorban}
 \ead{ag153@le.ac.uk}
\address{Department of Mathematics, University of Leicester, Leicester, LE1 7RH, UK}
\author{G.~S.~Yablonsky}
 \ead{gy@seas.wustl.edu}
\address{Parks College, Department of Chemistry, Saint Louis
University, Saint Louis, MO 63103, USA}

%\maketitle

\end{frontmatter}
The century of complexity has come. Many people write and speak about complexity. The
statement of the great physicist Stephen Hawking, ``I think the next century will be the
century of complexity,'' in his `millennium' interview on January 23, 2000 (San Jose
Mercury News) became a widely cited prophecy.

\begin{figure}[h]
\centering{
\fbox{\includegraphics[width=0.4\textwidth]{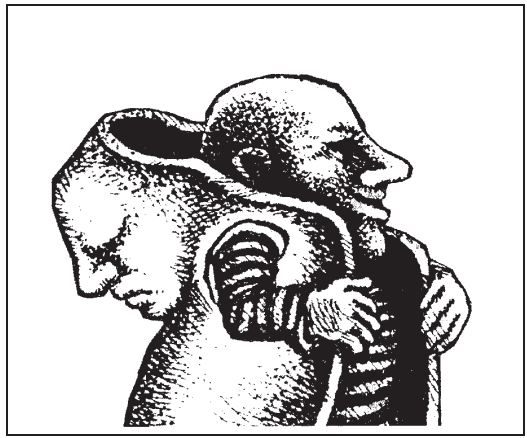}}
\caption{\label{Fig:Changes}Change of era: The direction  is changed
dramatically and the history of our motion is like a hood behind our shoulders.
To describe our recent direction we need to understand our past. Graphics by {\em Mikhail Molibog}.}}
\end{figure}

The face of science has changed (see cartoon in Fig.~\ref{Fig:Changes}). Surprisingly, when we start asking about the essence of
these changes and then critically analyze the answers, the result are mostly
discouraging. Why do we talk about complexity? Somebody might answer that now we have to study
{\em non-linear} systems and therefore they are complex. The answer seems to be plausible,
nonlinearity results in non-additivity of parts and in the emergence of new phenomena: ``The
whole is more than the sum of its parts.'' But objection appears immediately: non-linearity
has been in the focus of scientific research already for more than a century. Poincar\'e and
Lyapunov have studied nonlinear systems more than a century ago. Boltzmann's equation
and Navier--Stokes equation, the great nonlinear equations are more than a century old.
Many ideas have been created and many methods developed. The study of non-linearity is
not a symptom of the change of era. More than a thousand years ago Aristotle had written that
``the whole is something besides the parts'' (Metaphysics, Book 8, Chapter 6) and the
Western culture had accepted this idea from the very beginning. By the way, `besides' in
this translation of Aristotle sounds much more precise than the widely spread `more'.

We need another idea to understand the recent change of era and some people add that we
have to study large systems, both large and non-linear. Does the idea of {\em large
dimension} give us the key for understanding of new era? Not precisely! The curse of
dimensionality is now a well known problem and the term was proposed by Bellman in 1950s
\cite{Bellman1961}. Fifty years before, in 1900, David Hilbert in his address to  the
International Congress of Mathematicians in Paris has described 23 major mathematical
problems to be studied in the coming century \cite{Hilbert1900}. The title of one of
these problems sounds very strange and too broad ``Mathematical treatment of the axioms
of physics'' but if we read beyond the title then we immediately realize what has been
the main problem for Hilbert: ``As to the axioms of the theory of probabilities, it seems
to me desirable that their logical investigation should be accompanied by a rigorous and
satisfactory development of the method of mean values in mathematical physics, and in
particular in the kinetic theory of gases.'' He continues: ``Boltzmann's work on the
principles of mechanics suggests the problem of developing mathematically the limiting
processes, there merely indicated, which lead from the atomistic view to the laws of
motion of continua.'' In the modern scientific jargon, Hilbert had asked about the
correct methods of level jumping and model reduction, from large number of interacting
particles to mechanics of continua. For this purpose, he proposed to develop the theory
of probability and other related disciplines. This is the struggle with complexity of
large nonlinear systems recognized as one of the most important problems for mathematics
of the 20th century.

It is normal when the change of epochs under close examination looks as a continuous
development, not as a jump. But we talk about a century of complexity and suddenly find
that it was started more than a century ago. Perhaps, the idea `nonlinearity+large
dimension' cannot separate the new era  in spite of its attractiveness and clearness. To
understand the essence of changes we have to ask not only what appears but also what has
gone (Fig.~\ref{Fig:Changes}).

\begin{figure}
\centering{
\includegraphics[width=0.6\textwidth]{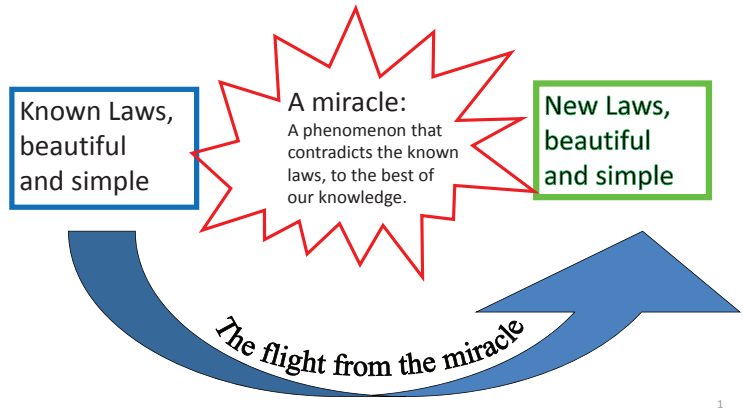}
\caption{\label{Fig:Flight}The flight from miracle: Einstein's road.}}
\end{figure}

What have been the most important scientific achievements of the 20th century? The new
great laws: the  great parade of the great discoveries, from the relativity and quantum mechanics
to genetics and DNA. One of the main players of this great period, Albert Einstein, has
described the discovery of the new laws as a ``flight from miracle'': ``The development
of this world of thought is in a certain sense  a continuous flight from the `miracle'." [``Die Entwicklung
dieser Gedankenwelt ist in gewissem Sinn eine best\"andige Flucht aus dem `Wunder','' \cite{EinstAutoBio}.] What
does it mean? Let us imagine: we have the laws, beautiful and simple (the Newton
mechanics, for example). We find a phenomenon that we cannot describe using these laws.
This is a {\em miracle}, a phenomenon that contradicts the basis laws. We trust in these laws,
we know that they are supported by the previous development of science, we like them
and try to use them again and again to describe the miracle. If we fail then we have to
use another way. We like our laws but we like the rationality more, therefore we fly from
the miracle by inventing new laws, which are beautiful, simple and, at the same time,
allow us to describe the phenomenon. After that, the miracle disappears and we have new
laws, beautiful and simple (Fig.~\ref{Fig:Flight})

This scheme can be explained much deeper with more historical details and examples, but the
main steps are clear: we look for a miracle and find a phenomenon that seems to be
in contradiction with the basic laws; we try to demystify this miracle by rational
explanations and models based on these laws; after several attempts and failings we decide that new
laws are needed and try to find new beautiful and simple laws that demystify the new phenomenon and
still can explain the other known phenomena not worse than the old laws.

A new scheme of actions became dominant in the struggle with complexity. The complexity
is recognized as the gap between the laws and the phenomena. We assume that the laws are
true. We can imagine a `detailed' model for a phenomenon but because of complexity, we
cannot work with this detailed model. For example, we can write the Schr\"odinger
equation for nuclei and electrons (formally, using indexes and signs of summation) but we
cannot use them directly for modeling of materials or large molecules. We can imagine a
detailed kinetic equation for a reaction network but cannot find reaction rate constants
and cannot work with this large system even if it is true.

In some cases, bridging this gap between the laws and the phenomena can be achieved in model engineering by
the special interaction between theoretical and experimental  studies, and real engineering as well.
Both the basic theory and the experiment will support the process of modelling. They may substitute for each other.
For example, we can make experiments instead of solving the extremely complicated equations.
We are sure that the answer should be the same after filtering the noise for experimental errors.
We also can organise computational experiments instead of real ones. Again, we are sure that the answer should be the same after
cleaning the results from the errors. In the background of this belief the fundamental assumption that the possible world of the theory
coincides  with the real world of our experiments and practice (with sufficient accuracy).
We can believe that somewhere else, for high energies, very small distances, or very large distances
we need new laws, but not now.

The interaction between theory and experiment in the model engineering may generate not only mathematical models but
new experimental technics as well. For example, in chemistry, non-steady-state activity
screening can be based on the technique of Temporal Analysis of Products (TAP),
invented by John Gleaves in 1988. The main idea of TAP is to treat the catalyst
by a series of pulses of very small intensity relative to the amount of catalyst \cite{GleavesYab1997}.
This infinitesimal approach can be termed `chemical calculus'.

\begin{figure}
\centering{
\includegraphics[width=0.7\textwidth]{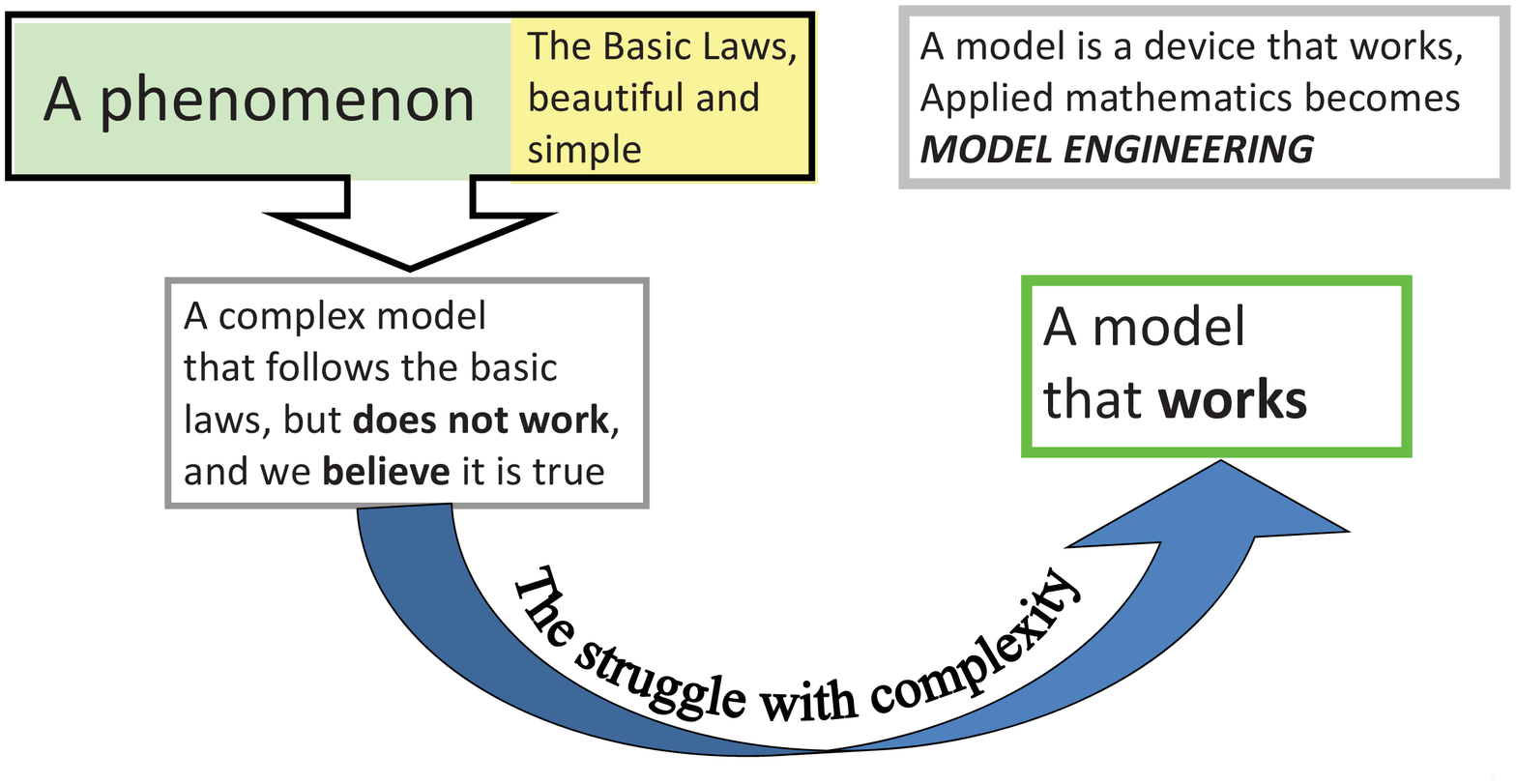}
\caption{\label{Fig:Struggle}Struggle with complexity: the life battle
 of the model engineers.}}
\end{figure}

The result of the struggle with complexity is a model that works. This is a sort of
engineering: a model is a device and this device should be functional. Applied
mathematics and mathematical modelling become a sort of engineering and instead of
Einstein's flight from miracle (Fig.~\ref{Fig:Flight}) another scheme arises
(Fig.~\ref{Fig:Struggle}): We know the laws and we have a phenomenon. We need a model for
work. For different work need different models are needed. We may combine the first
principles, the empirical data and even the active experiment to create the model. There
exist special technologies for testing and validation of models. The structure of the
whole process seems to be similar to the design of machines and it might be reasonable to
teach the students in applied mathematics the module of `Systems Engineering', as a guide
the engineering of complex systems \cite{SysEng2011}.

The focus has moved from the revolution in laws to the  production of intellectual devices. In the context of the natural sciences
this is model making under given basic laws. On  the other hand, the systems under consideration may be artificial and instead of the basic laws we
deal with the man-made plans, projects and scenarios. Such systems as the Internet, social institutions, large plants,
financial system and many other systems are now in the focus of attention together with natural phenomena.
The hybrid systems, that obey the natural laws but experience significant influence of human activity and man-made projects
are of great interest too, like climate or biosphere.

The nature does not change and there will be many new laws to discover. The application of science always exists too.
The era of complexity is in the change of the {\em focus} of the research activity.
From the epoch of the great scientific revolutions we have moved to the epoch of the
intellectual devices, from the revealing the God's or Nature plan to the intellectual
engineering at various scales that is necessary to provide tools for prediction of the results of human activity.
The new epoch may be ended some day but this is difficult to predict.

The milestones of development rarely coincide with the ends of calendar centuries. We
believe that the `century of scientific revolutions' is situated between two giants, from
L. Boltzmann to R.P. Feynman. Surprisingly, their contribution in the era of complexity
is also huge. We can just recall Boltzmann's entropy \cite{Boltzmann1872} and Feynman's
inventions of nanotechnology \cite{Feynman1960} or quantum computers \cite{Feynman1982}.

In the struggle with complexity there are many specific problems and tools. This issue
presents several slices of this activity:
\begin{enumerate}
\item{Measuring complexity: the curses and blessings of dimensionality;}
\item{Model reduction and invariant manifolds;}
\item{Fingerprinting, criteria, and interpretation of experiments;}
\item{Modelling of classes of complex systems.}
\end{enumerate}

In the first part, {\em Measuring complexity: the
curses and blessings of dimensionality}, the general problems are approached. The first
general problem is the {\em curse of dimensionality}. V. Pestov \cite{Pestov2013}
demonstrates how the curse of dimensionality affects the nearest neighbor search and the
widely used kNN classifiers. He demonstrates how the performance of the kNN classifier in
very high dimensions can become unstable. Then, he develops a procedure for the reduction of
the multidimensional statistical learning problems to a one-dimensional problem by a
Borel isomorphism of the spaces with measure.

High dimensional problems are not always complex. From a certain point of view, they look much
simpler: the central limit theorem in probability and the advanced results about {\em
measure concentration} \cite{GianMilman2000,Talagand1996,Gromov2007} demonstrate how
convex sets in high dimension become `almost spheres', and typical distribution functions
look like Gaussians. This phenomenon (we call it the {\em blessing of
dimensionality}) was recognised first in statistical physics by Maxwell and Gibbs
\cite{Gibbs1902}. For multiparticle systems (under some technical assumptions) the
microcanonical ensemble with the given values of energy is equivalent to the canonical one
which can be represented by the entropy maximum with the same average energy. The Maximum
of Entropy (MaxEnt) approach naturally appears in the limit of high dimension.

In the middle of the 20th century, after C. Shannon's works \cite{Shannon1948} and E.T.
Jaynes papers \cite{Jaynes1957}, the MaxEnt approach became very popular as a
maximization of the subjective uncertainty measured by the Boltzmann--Gibbs--Shannon
entropy. In 1960, A. R\'enyi invented non-classical entropies \cite{Renyi1961}.
Csisz\'ar, Morimoto, Tsallis and many other researchers developed this idea further,
and now we have the rich choice of the entropies for many problems. This rich choice
leads to the `uncertainty of uncertainty problem': which entropy to use for the
uncertainty measurement? A.N. Gorban proposes to use all the entropies together
\cite{Gorban2013}. This approach results in a set of conditionally "most random"
distributions. Surprisingly, this set allows constructive description. This new
`Maxallent' (Maximizers of all Entropies) method is based on the understanding of entropy as
a measure of uncertainty which increases in Markov processes \cite{GorGorJudge2010}.

In the work of M. Grmela \cite{Grmela2013}, the Dynamical Maximum Entropy Principle is
elaborated. It covers equilibrium and non-equilibrium thermodynamics and gives new
approaches to some classical problems. In particular, the classical Chapman--Enskog
expansion in the theory of Boltzmann's equation \cite{ChapmanCowling1970} is described by
the entropy deformation.

A. Zinovyev and E. Mirkes develop the data approximation approach to measure the
complexity of datasets \cite{ZinovyevMirkes2013}. They utilize the universal
approximators, principal cubic complexes, and generalize the notion of principal
manifolds and graphs \cite{LNCSE58} for datasets with nontrivial topologies and are
constructed with a grammar of elementary graph transformations. Three natural types of
data complexity are used and tested in the case studies: the geometric,  structural and
construction complexity.

Idempotent and tropical mathematics provide asymptotic versions of the classical
mathematics produced by the `dequantization' procedure \cite{LitvinovMaslov2005}. G.L.
Litvinov evaluates the complexity of the algorithms for the idempotent problems and their
interval versions and demonstrated that they may be much simpler than in the classical
mathematics \cite{Litvinov2013}.

Model reduction is one of the major procedures in the struggle with complexity and
section {\em Model reduction and invariant manifolds} in the issue includes papers
about reduction of dynamical models. M. Slemrod \cite{Slemrod2013} revisits the sixth
Hilbert problem and demonstrates that the solution has to be negative for compressible
gas dynamics: the hydrodynamic limit does not lead to the classical compressible Euler or
Navier--Stokes equations. This situation differs from the incompressible limit
\cite{Saint-Raymond2009}. The key to this analysis is provided by the exactly solvable
reduction models discovered by A.N. Gorban and I.V. Karlin
\cite{GorKarPRL1996,KarGorAnn2002}.

Slow invariant manifolds are the main tools for model reduction in dissipative systems
\cite{ConstTemam1988,GorKarLNP2005}. The fast manifold traditionally attracts less
attention and plays an auxiliary role. It is used mostly for projection of a motion on an
approximate invariant manifold. V. Bykov and V. Gol'dshtein \cite{BykovGol2013}
demonstrate how to start model reduction procedures from fast manifolds and develop a
theory of Singularly Perturbed Vector Fields (SPVF) with the main emphasis on fast
invariant manifolds. The slow manifold appears as a by-product of this approach. The new
approach is illustrated by the examples from chemical kinetics.

The Lam and Gousis Computational Singular Perturbation (CSP) approach aims to find both
fast and slow manifolds for a system of differential equations \cite{Lam1994}. It was
developed for application in chemical kinetics. In their paper \cite{Gousis2013}, P.D.
Kourdis, A.G. Palasantz, and D.A. Goussis develop the algorithmic realization of CSP and
apply it to important biochemical systems with oscillations, the NF-$\kappa$B signaling system.

The problem of model reduction for systems with symmetries is analyzed by B. Sonday, A.
Singer and I.G. Kevrekidis \cite{Kevrekidis2013}. They use the Kuramoto-Sivashinsky
equation with periodic boundary conditions and a stochastic simulation of nematic liquid
crystals as examples, and apply the eigenvector-based techniques for model reduction. They
also use a new technic, Vector Diffusion Maps \cite{Singer2012}, that combines, in a
single formulation, the symmetry removal step and the dimensionality reduction step.

B.R. Noack, R.K. Niven \cite{NoackNives2013} develop further a MaxEnt closure strategy for  
Galerkin systems arising from a projection of the incompressible Navier-Stokes equation onto orthonormal expansion modes. 
They aim to discover and demonstrate a new face of the turbulence closure problem.

R. Hannemann-Tamas, A. Gabor, G. Szederkenyi, and K.M. Hangos formulate the model
reduction problem for chemical kinetics as a quadratic programming problem
\cite{Hangos2013}. The objective function is derived from the parametric sensitivity
matrix. The method eliminates unnecessary reactions for a given level of tolerance and
adjusts the rate constants of the remaining reactions for error minimization. The
efficiency of the approach is demonstrated on the known benchmarks.

The transition from dynamics to thermodynamics is the most complicated step on the stair of
reduction \cite{GorKarLNP2005}. In the paper by T. Chumley, S. Cook, and R. Feres
\cite{Feres2013} this step is analyzed for billiard-like random systems. These systems
exhibit irreversible thermodynamics behavior, indeed.

The ideal model reduction technology starts from the detailed system and produces the
reduced one. This picture may be oversimplified. Indeed, in many practically important
cases the mathematical model cannot be produced without simplifications and model
reduction becomes a tool for model construction from scratch. It may be also used for
construction of semi-empirical methods and active theory-driven experiments. In
engineering, many semi-empirical criteria were invented to separate regimes: laminar from
turbulent, shocks from smooth incompressible flows and many others.  The modern
fingerprinting idea may find its logical roots in the semi-empirical criteria. ``The goal
of the fingerprint analysis is to find features and characteristics of observed complex
behavior, based on which it is possible to find out the model, its class or its family,
and to determine its characteristics'' \cite{MarYab2011}. The fingerprints, patterns, signatures or
motifs allow us to work with complex systems without extraction of deep and expensive
information. Kinetic signatures in biochemical reactions  \cite{MorRNA2012}, motifs of genetic
sequences \cite{GeneMotif2003} patterns in time series \cite{PatternsCardio}
(cardiogramms and encephalogramms, for
example) give us nice examples of fingeprinting.

The paper by D. Constales, G.S. Yablonsky, and G.B. Marin \cite{ConstYabMar2013} opens Section {\em
Fingerprinting, criteria, and interpretation of experiments}. They study the basic
patterns in simple reaction networks. This work aims to analyze appearance of some basic
patterns in chemical kinetics, to review and extend the previous findings \cite{YabConst2010}.
Authors supplement the classical notion of complexity by
`simplexity' to reflect the rich diversity of  patterns which can be produced even by
simple systems.

A useful example of a criterion validation is given in the work by F. Xia and R.L. Axelbaum
\cite{XiaAxelbaum2013}. They propose to use the local ratio C/O to classify various regimes and zones of
diffusion flames. Radical pool and soot precursor zones are shown to be clearly
delineated in C/O ratio space. This ratio is validated as a criterion for interpreting flame
structure.

M.J. Hankins, T. Nagy, and I.Z. Kiss \cite{HankinsKiss2013} develop an original technology for active
experiment for construction of nullcline-based models and demonstrates its efficiency on the modelling of
the electrochemical reaction. Perhaps, the first author who proposed to use the
nullcline-based models instead of detailed differential equations was A.N. Kolmogorov
\cite{Kolmogorov1936,SigmundKolmogorov2007}. M.J. Hankins et al use the nullcline-based models with the
singular pertirbation assumption (time scale separation). Under this assumption, the
nullclines may be extracted from the control experiment with a combination of active and
proportional controllers acting on the  fast and the slow variables.

The section {\em Modelling of classes of complex systems} includes four papers about four
classes of systems: networks, finance, catalysis (in chemical engineering) and
bioreactors. The new tools and case studies are presented. H. Sayama, I. Pestov,
J. Schmidt, B.J. Bush, C. Wong, J. Yamanoi, and T. Gross describe the methods based on
adaptive networks with self-organization of structure for modeling of complex networks
like social, transportation, neural and biological networks \cite{SayamaPestov2013}.  B.E. Baaquie describes
a quantum mathematics approach to financial modelling \cite{Baaquie2013}.
F.J. Keil presents a thorough review about modelling in catalysis, from quantum chemical methods for calculating
reactions on the active centers to transport in porous media \cite{Keil2013}. I. Iliuta and F. Larachi
study dynamics of bacterial cells in trickle-bed bioreactors. They model the basic
processes, fluxes in multiface flows, population balance for cells and agglomerates,
biomass dynamics, dynamics of agglomeration and filtration \cite{IliutaLarachi}.

Neither one issue of a journal,  nor a large encyclopedia can capture everything about such a
broad and dynamic subject as grasping complexity, but we hope that various  faces of the
modern era of complexity are presented here.


\begin{thebibliography}{99}
\bibitem{Pestov2013}V. Pestov, Is the k-NN classifier in high dimensions affected by the
    curse of dimensionality? Computers and Mathematics with Applications, This issue.
\end{thebibliography}
\end{document}